\documentclass[prl,aps,superscriptaddress,twocolumn]{revtex4}
\usepackage{latexsym}
\usepackage{amsfonts}
\usepackage{amssymb}
\usepackage{amsmath}
\usepackage{graphicx}
\begin{document}
\title{Evidence of Doping-Dependent Pairing Symmetry in
Cuprate Superconductors}
\date{\today}
\author{N.-C. Yeh}
\author{C.-T. Chen}

\affiliation{Department of Physics, California Institute of
Technology, Pasadena, CA 91125}

\author{G. Hammerl}
\author{J. Mannhart}
\author{A. Schmehl}
\author{C. W. Schneider}
\author{R. R. Schulz}

\affiliation{Center for Electronic Correlations and Magnetism, Institute of
Physics, Augsburg University, D-86135 Augsburg, Germany}

\author{S. Tajima}
\author{K. Yoshida}

\affiliation{Superconductivity Research Laboratory, International Superconductivity Technology Center, 1-10-13, Sinonome, Koto-ku, Tokyo 135, Japan}

\author{D. Garrigus}
\author{M. Strasik}

\affiliation{Boeing Information, Space \& Defense Systems, Renton, WA 98055}

\begin{abstract}
Scanning tunneling spectroscopy (STS) studies reveal long-range
spatial homogeneity and predominantly $d_{x^2-y^2}$-pairing
spectral characteristics in under- and optimally doped $\rm
YBa_2Cu_3O_{7-\delta}$ superconductors, whereas STS on $\rm
YBa_2(Cu_{0.9934}Zn_{0.0026}Mg_{0.004})_3O_{6.9}$ exhibits {\it
microscopic} spatial modulations and strong scattering near the Zn
or Mg impurity sites, together with global suppression of the
pairing potential. In contrast, in overdoped $\rm
(Y_{0.7}Ca_{0.3})Ba_2Cu_3O_{7-\delta}$, $(d_{x^2-y^2}+s)$-pairing
symmetry is found suggesting significant changes in the
superconducting ground-state at a critical doping value.
\end{abstract}
 \maketitle

One of the most important issues associated with the cuprate
superconductors is the possible existence of a quantum critical
point (QCP) near the optimal hole doping ($p_o \approx 0.16$ per
$\rm CuO_2$) in these doped Mott antiferromagnets
\cite{Varma97,Vojta00,Chakravarty00}. Although non-monotonic
doping-dependent physical properties in the underdoped ($p < p_o$)
and overdoped ($p > p_o$) regimes are suggestive of the presence
of a QCP, an unambiguous proof would require identifying the
relevant broken symmetry associated with a critical doping level
$p_c$. Various theoretical studies\cite{Vojta00,Lee00} have
investigated the dependence of the ground state of doped Mott
antiferromagnets on the doping level and the strength of exchange
and Coulomb interactions \cite{Vojta00}. The relevant symmetries
associated with the competing orders include: electromagnetic
$U(1)$, spin rotation-invariance $SU(2)$, square lattice space
group $C_{4v}$ and time-reversal symmetry ${\cal T}$
\cite{Vojta00}. Experimentally, while the $d_{x^2-y^2}$ pairing
symmetry is known to dominate in the optimally doped cuprates
\cite{Tsuei00,VanHarlingen95,Yeh01,Yeh00,Wei98}, possible
doping-dependent pairing symmetry has been suggested
\cite{Deutscher99}. In general, the doping dependence of the
pairing symmetry and the issue of quantum criticality must be
considered under the premise of spatial homogeneity in the pairing
potential. Recent finding of nano-scale variations in the measured
energy gap of $\rm Bi_2Sr_2CaCu_2O_{8+x}$ (Bi-2212) \cite{Pan01}
and the implication of macroscopic clusters in overdoped $\rm
La_{2-x}Sr_xCuO_4$ \cite{Wen00} present possible complications in
unraveling the existence of a QCP.

A possible consequence of a QCP is the doping-dependent pseudogap
phenomenon \cite{Timusk99}, which may represent a precursor for
superconductivity in the cuprates \cite{Timusk99}. Early
experiments on the Bi-2212 system reported a measured energy
gap $\Delta ^{\ast}(p)$ that increased monotonically with
decreasing $p$ and was nearly independent of temperature
\cite{Miyakawa99,Miyakawa98,Renner98,Norman98,Fedorov99,Shen95}.
However, the low-temperature spectra of the optimally doped and
underdoped Bi-2212 appeared to consist of a sharp peak feature
on top of a broad ``hump''. Recent bulk measurements on Bi-2212
mesas \cite{Krasnov00} demonstrated strong temperature
dependence associated with the sharp peak, which vanished at the
superconducting transition temperature $T_c$, while the hump
feature persisted well above $T_c$. The coexistence of these two
gap-like features in the superconducting state has been attributed
to different physical origin associated with each gap
\cite{Krasnov00,Bang00}.

In this work, we address some of these issues via studies of the
directional and spatially resolved quasiparticle tunneling spectra
on the $\rm YBa_2Cu_3O_{7-\delta}$ (YBCO) with a range of doping
levels. The doping dependence of the pairing symmetry, pairing
potential and spatial homogeneity is derived from these studies.

The samples used in this investigation included three optimally
doped YBCO single crystals with $T_c = 92.9 \pm 0.1$ K, three
underdoped YBCO single crystals with $T_c = 60.0 \pm 1.5$ K, one
underdoped YBCO c-axis film with $T_c = 85.0 \pm 1.0$ K, two
overdoped $\rm (Y_{0.7}Ca_{0.3})Ba_2Cu_3O_{7-\delta}$ (Ca-YBCO)
c-axis films \cite{Schneider99} with $T_c = 78.0 \pm 2.0$ K, and
one optimally doped single crystal containing small concentrations
of non-magnetic impurities, $\rm
YBa_2(Cu_{0.9934}Zn_{0.0026}Mg_{0.004})_3O_{6.9}$ ((Zn,Mg)-YBCO),
with $T_c = 82.0 \pm 1.5$ K \cite{Yeh00,Schneider99}. The spectra
of YBCO single crystals were taken primarily with the
quasiparticles tunneling along three axes: the antinode axes
$\{100\}$ or $\{010\}$, the nodal axis $\{110\}$, and the c-axis
$\{001\}$; while those of the pure and Ca-YBCO films were taken
along the c-axis. All samples except (Zn,Mg)-YBCO are twinned. The
surface was prepared by chemical
etching\cite{Vasquez89a,Vasquez89b}, and samples were kept either
in high-purity helium gas or under high vacuum at all times. Our
surface preparation has the advantage of terminating the YBCO top
surface at the $\rm CuO_2$ plane by chemically passivating the
layer while retaining the bulk properties of the constituent
elements \cite{Vasquez89a,Vasquez89b}, thus yielding reproducible
spectra for samples of the same bulk stoichiometry, although
direct constant-current mode atomic imaging of the chemically
inert surface becomes difficult. For comparison, surfaces of
vacuum-cleaved YBCO samples are found to terminate at the
CuO-chain layer, which are prone to loss of oxygen and development
of surface states \cite{Vasquez89a,Vasquez89b}. For vacuum-cleaved
Bi-2212, the surface typically terminates at the BiO layer.

\begin{figure}
  \centering
  \includegraphics[keepaspectratio=1,height=4.1in]{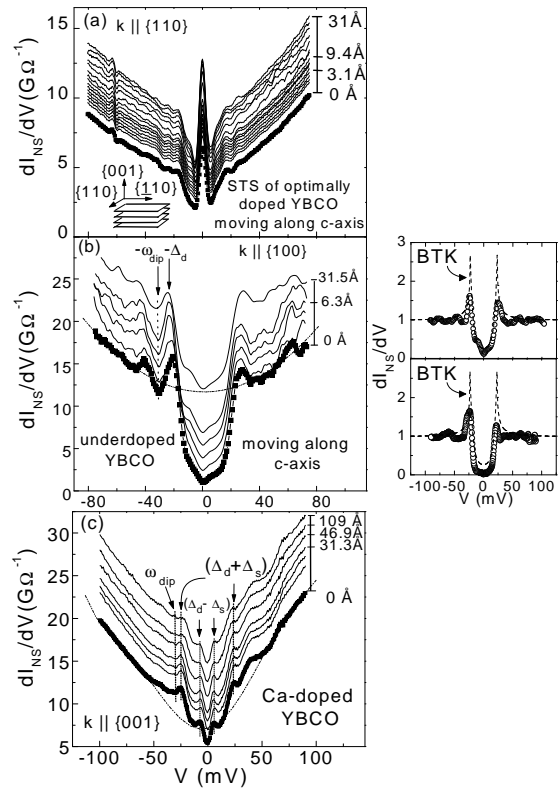}
\caption{Spatially resolved ($dI_{NS}/dV$) vs. ($V$) spectra at
4.2 K: {\bf (a)} Optimally doped YBCO. {\bf (b)} Underdoped YBCO.
Panels: normalized $\{100\}$ spectra of an underdoped YBCO (upper
panel) and (Zn,Mg)-YBCO (lower panel) together with a BTK fitting
curve (dashed line). {\bf (c)} Ca-YBCO. The tunneling cone
\cite{Yeh01,Yeh00,Wei98} of all spectra ranges from $(\pi/12)$ to
$(\pi/8)$. For the BTK analysis, each set of the data was
normalized relative to the polynomial fit to the high-voltage
background conductance, as shown in {\bf (b)} and {\bf (c)} by the
dashed-dotted curve.}\label{fig1}
\end{figure}

\begin{figure}
  \centering
  \includegraphics[keepaspectratio=1,width=3.36in]{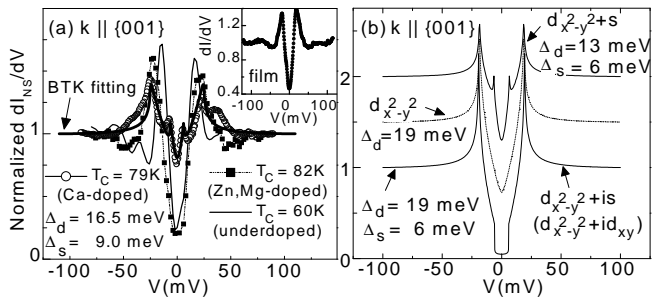}
\caption{{\bf (a)} Normalized c-axis tunneling spectra of one
underdoped, one (Zn,Mg)-doped, and one Ca-YBCO, together with a
BTK fitting curve to the Ca-YBCO spectrum (the thick solid line).
Inset: c-axis spectrum of an underdoped YBCO film. {\bf (b)}
Calculated spectra for different pairing symmetries.}\label{fig2}
\end{figure}

Figure \ref{fig1} illustrates representative tunneling conductance
$(dI_{NS}/dV)$ vs. voltage $(V)$ data with high spatial resolution
for YBCO samples at 4.2 K: (a) optimally doped YBCO, with the
average quasiparticle momentum $\vec k \parallel \{110\}$ and the
tip scanning along $\{001\}$; (b) underdoped YBCO, with $\vec k
\parallel \{100\}$ and scanning along $\{001\}$; and
(c) Ca-YBCO, with $\vec k \parallel \{001\}$ and scanning along
$\{100\}$. The normalized $(dI_{NS}/dV)$ vs. $V$ spectra of under-
and (Zn,Mg)-doped YBCO for $\vec k \parallel \{100\}$ are shown in
the inset of Fig. \ref{fig1}(b), and those of underdoped,
(Zn,Mg)-doped, and Ca-doped YBCO for $\vec k \parallel \{001\}$
are illustrated in Fig. \ref{fig2}(a). To verify the spatial
resolution, we also performed $(dI_{NS}/dV)$ mapping of an
optimally doped single crystal at a constant bias voltage of $V
\approx (\Delta _d/e)$ with tip scanning along $\{001\}$ and $\vec
k \parallel \{100\}$. Periodic conductance modulations consistent
with the known atomic layer separations in YBCO were resolved.
Independent imaging on NbSe$_2$ also confirmed atomic spatial
resolution.

For the optimally doped and underdoped YBCO, the STS exhibited
long-range ($\sim 100$ nm) spatial homogeneity and strong
directionality, showing a zero-bias conductance peak (ZBCP) for
$\vec k \parallel \{110\}$ \cite{Yeh01,Yeh00,Wei98,Hu94,Tanaka95},
(Fig. \ref{fig1}(a)); nearly ``U-shape'' gap features around the
zero bias for $\vec k \parallel \{100\}$, (Fig. \ref{fig1}(b));
and ``V-shape'' features for $\vec k \parallel \{001\}$, (Fig.
\ref{fig2}(a)). By taking into account a finite transverse
momentum distribution for the incident quasiparticles relative to
the normal of the crystalline plane, (i.e., by considering a
finite ``tunneling cone'', parameterized as $\beta$ with a typical
value ranging from $15^{\circ}$ to $22.5^{\circ}$ \cite{Wei98}),
the primary features and the directionality of the spectra of all
optimally doped and underdoped YBCO samples were consistent with
the generalized Blonder-Tinkham-Klapwijk (BTK) theory by Hu
\cite{Hu94}, Kashiwaya and Tanaka \cite{Tanaka95} for
$d_{x^2-y^2}$ pairing \cite{Yeh01,Yeh00,Wei98,Hu94,Tanaka95}.

For the Ca-doped YBCO epitaxial films, {\it macroscopic} spatial
variation in the STS at a length scale of $\sim 50$ nm was
observed, which correlated with the dimension of the growth
islands according to images of atomic force microscopy
\cite{Schneider99}, while the STS within each island were
spatially homogeneous, as exemplified in Fig. \ref{fig1}(c). The
spatial modulation of STS may be the result of macroscopic phase
segregation, similar to the finding in overdoped LSCO
\cite{Wen00}. Furthermore, the STS of Ca-YBCO exhibited long-range
and symmetric ``subgap peaks'', (see Figs.
\ref{fig1}(c) and \ref{fig2}(a)), which differed from the spectral
contributions of local disorder (such as oxygen vacancies), the
latter generally appeared as short-range and asymmetric ``humps'' in
the spectra. These spectra were in good agreement with the
($d_{x^2-y^2}+s$) pairing symmetry according to the generalized
BTK analysis \cite{Yeh01,Yeh00,Wei98,Hu94,Tanaka95}, with a
pairing potential $\Delta _k = \Delta _d \cos 2 \theta _k + \Delta
_s$, where $\theta _k$ is the angle of $\vec k$ relative to $\{100\}$.
We obtained two typical sets of spectra with $\Delta _d = 17$ meV
and $\Delta _s = 9$ meV for data in Fig. \ref{fig1}(c),
and $\Delta _d = 13$ meV and $\Delta _s = 6$ meV
for the other set of spectra given elsewhere \cite{Yeh01}. On the
other hand, $d_{x^2-y^2}$ pairing symmetry was confirmed on
the optimally doped and underdoped YBCO c-axis epitaxial films
prepared by two research groups, as exemplified in the inset of
Fig. \ref{fig2}(a) for a YBCO film with $T_c = 85.0
\pm 1.0$ K. For clarity, we depict the calculated c-axis
tunneling spectra in Fig. \ref{fig2}(b). Thus, under the premise
that the spectra are spatially homogeneous well beyond the
coherence length and mean free path, we suggest that {\it the
long-range and symmetric subgap peaks in overdoped YBCO represent
supporting evidence for doping-induced variations in the pairing
symmetry.} The substantial $s$-pairing component ($> 20\%$) may
also contribute to the significant enhancement in the critical
current of bi-crystal Josephson junctions and polycrystalline
samples based on the Ca-YBCO \cite{Schneider99}.

\begin{figure}
  \centering
  \includegraphics[keepaspectratio=1,height=2.88in]{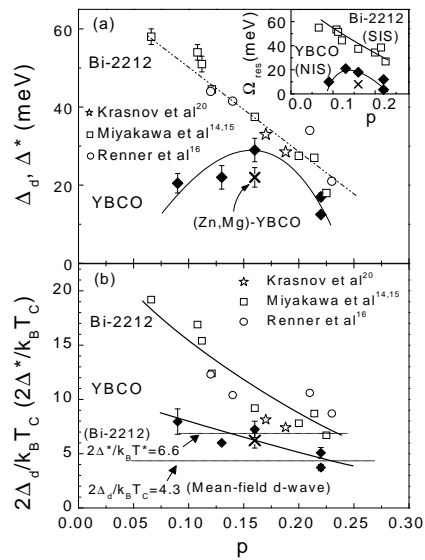}
\caption{{\bf (a)} Comparison of $\Delta _d (p)$ in YBCO with
$\Delta ^{\ast} (p)$ in Bi-2212. The doping level $p$ is
determined from the formula $T_c = T_{c,max} \lbrack 1 -
82.6(p-0.16)^2 \rbrack$ \cite{Yeh01} with $T_{c,max} = 93.0$ K.
Inset: Comparison of $\Omega _{res} (p)$ for YBCO and Bi-2212.
{\bf (b)} Doping-dependent ($2 \Delta _d/k_B T_c$) for YBCO and
($2 \Delta ^{\ast}/k_B T_c$) for Bi-2212. The error bar associated
with each doping level of YBCO covers the range of $\Delta _d$ (or
$\Omega _{res}$) obtained from all spectra and the uncertainties
of the BTK fitting.}\label{fig3}
\end{figure}

The maximum value of the $d$-wave gap, $\Delta _d$, appeared to be
non-monotonic with the doping level $p$ (Fig. \ref{fig3}(a)),
whereas the ratio $(2 \Delta _d/ k_B T_c)$ increased with
decreasing doping, from $\sim 7.8$ for $p \approx 0.09$ to $\sim
4.5$ for $p \approx 0.22$, as illustrated in Fig. \ref{fig3}(b).
On the other hand, experiments on Bi-2212 found an averaged gap
$\Delta ^{\ast}$ that scaled with the pseudogap temperature
$T^{\ast}$, $2 \Delta ^{\ast} \sim 6.6 k_B T^{\ast}$
\cite{Miyakawa99,Miyakawa98,Renner98,Norman98,Fedorov99,Shen95},
and that $\Delta ^{\ast}$ increased with decreasing $p$
\cite{Miyakawa99,Miyakawa98,Renner98,Norman98,Fedorov99,Shen95}.
We speculate that the superlattice and charge-density-wave
modulations of the BiO layers in Bi-2212 may have complicated the
quasiparticle spectra
\cite{Miyakawa99,Miyakawa98,Renner98,Norman98,Fedorov99,Shen95,Krasnov00,Bang00}.
Furthermore, the tendency of random oxygen distribution in
Bi-2212, as opposed to the generally ordered oxygen distribution
in YBCO, may be responsible for the nano-scale variations in
$\Delta ^{\ast}$ of Bi-2212, ranging from $\sim 15$ to $\sim 70$
meV over a few nanometers \cite{Pan01}.

Although the generalized BTK analysis was suitable for deriving
the pairing potential and the primary spectral characteristics, it
could not account for the ``satellite features'' associated with
many-body interactions \cite{Wu01} in all YBCO spectra. In
Bi-2212, the spectral ``dip/hump'' features were widely observed
\cite{Miyakawa99,Miyakawa98,Renner98,Norman98,Fedorov99,Shen95}
and attributed to quasiparticle damping via interactions with
collective spin fluctuations \cite{Wu01,Chubukov00}. In the strong
coupling limit, the spectral dip is expected to appear at the
energy $\omega _{dip} = \Delta + \Omega _{res}$, where $\Delta$ is
the measured gap, and $\Omega _{res}$ is associated with the
resonance of collective spin excitations \cite{Wu01,Chubukov00}.
Defining $\Omega _{res}$ as the energy difference between the
primary peak ($\Delta$) and the dip ($\omega _{dip}$) in the
spectra \cite{Chubukov00}, as indicated by the solid arrows in
Figs. \ref{fig1}(b) and \ref{fig1}(c), we found that $\Omega
_{res}$ in YBCO decreased with decreasing $p$, as shown in the
inset of Fig. \ref{fig3}(a).

\begin{figure}
  \centering
  \includegraphics[keepaspectratio=1,height=1.8in]{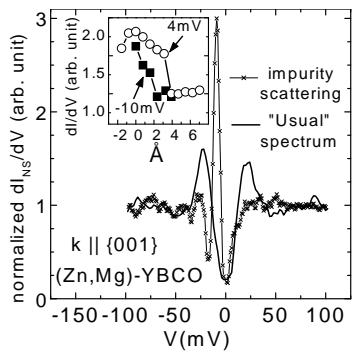}
\caption{$\{001\}$ tunneling spectrum of (Zn,Mg)-YBCO at a
spinless impurity site and a usual spectrum away from any
impurity. Two types of impurity scattering were observed: one
at $(-10 \pm 2)$ meV, the other at $(4 \pm 2)$ meV. Inset: Rapidly
decreasing intensity of the scattering peaks are shown for
displacement along $\{010\}$ from an impurity site.}\label{fig4}
\end{figure}

In the (Zn,Mg)-YBCO single crystal, {\it microscopic} spatial
variations were observed. Two types of impurity scattering spectra
were found. One was associated with a resonant scattering at ($-10
\pm 2$) meV, as shown in Fig. \ref{fig4}, the other at ($4 \pm 2$)
meV. Assuming the position where a local maximum intensity of a
resonant peak occurred as an impurity site, we found that the peak
persisted over several lattice constants for displacement along
either $\{100\}$ or $\{010\}$, and the peak intensity decreased
rapidly within the Fermi wavelength, as shown in the inset of Fig.
\ref{fig4}. Eventually the usual spectrum was recovered at about
two coherence lengths ($\sim 3$ nm) from an impurity site, similar
to Zn-doped Bi-2212 \cite{Pan00}. For displacement along other
directions, the spectral features became more complicated
\cite{Chen01}. Our data have confirmed that the spinless
impurities such as Zn$^{2+}$ and Mg$^{2+}$ are strong pair
breakers
\cite{Pan00,Chen01,Fong99,Figueras00,Balatsky95,Flatte98,Salkola98,Sidis00,Haas00,Polkovnikov01}.
Furthermore, the magnitude of the global $\Delta _d$ and $\Omega
_{res}$ in (Zn,Mg)-YBCO both became smaller than those of the
optimally doped YBCO, as shown in Fig. \ref{fig3}(a). The strong
suppression of $\Omega _{res}$ suggests that the spinless
impurity-induced moments may be responsible for the suppression of
collective spin excitations \cite{Polkovnikov01}. The single
resonance peak at the spinless impurity site is also in contrast
to the predicted double peaks (at $\pm E_0$) at the impurity site
had the ${\cal T}$-symmetry been broken due to a complex pairing
symmetry of either ($d_{x^2-y^2} + id_{xy}$) or ($d_{x^2-y^2} +
is$) \cite{Salkola98}.
In the context of quantum criticality, our data of
doping-dependent pairing symmetry is suggestive of a QCP with
broken $C_{4v}$-symmetry. However, a small $s$-pairing component
beyond the resolution of our STS might exist in the under- and
optimally doped YBCO due to the crystalline orthorhombicity, which
would imply that no obvious broken symmetry had taken place. For
comparison, no doping dependence in the pairing symmetry has ever
been observed in tetragonal Bi-2212. On the other hand, the
absence of either $(d_{x^2-y^2}+id_{xy})$ or $(d_{x^2-y^2}+is)$
pairing does not rule out the possibility of broken ${\cal
T}$-symmetry at a QCP, because certain broken ${\cal T}$-symmetry
states, such as the staggered flux \cite{Chakravarty00,Affleck88}
and the circulating current phase \cite{Varma97}, cannot be
detected directly by tunneling spectra. However, to date no
universal broken symmetry associated with a QCP can be identified
for all cuprate superconductors.

In summary, we have demonstrated long-range spatial homogeneity in
the quasiparticle tunneling spectra of YBCO samples with a range
of doping levels. In contrast, STS of the (Zn,Mg)-YBCO exhibit
strong pair-breaking effects near the Zn and Mg impurities. The
quasiparticle spectral characteristics and the impurity scattering
effects in optimally doped and underdoped YBCO are consistent with
predominantly $d_{x^2-y^2}$ ($> 95$\%) pairing symmetry, whereas
those of the overdoped Ca-YBCO exhibit $(d_{x^2-y^2}+s)$ pairing
symmetry with a significant $s$-component ($> 20$\%).

\begin{acknowledgments}
The work at Caltech was supported by NSF Grant \#DMR-9705171, at
University of Augsburg by BMBF Grant \#13N6918/1, and at SRL by
the New Energy and Industrial Technology Development Organization
as Collaborative Research and Development of Fundamental
Technologies for Superconductivity Applications. We thank
Professor Asle Sudb\o \ and Dr. C. C. Tsuei for useful comments
and stimulating discussions.
\end{acknowledgments}

\end{document}